\documentclass[prb,twocolumn,showpacs]{revtex4}

\usepackage{color}
\usepackage{bm}
\usepackage{amsmath}
\usepackage{graphicx}
\usepackage{amssymb}
\usepackage{bm}
\usepackage{times}

\begin{document}

\title{Deterministic preparation of Dicke states of donor nuclear spins in silicon by cooperative pumping}
\author{Yu Luo} \author{Hongyi Yu} \author{Wang Yao}
\thanks{wangyao@hkucc.hku.hk}
\affiliation{Department of Physics and Center of Theoretical and Computational Physics, The University of Hong Kong, Hong Kong, China}

\begin{abstract}
  For donor nuclear spins in silicon, we show how to deterministically
  prepare various symmetric and asymmetric Dicke states which span a
  complete basis of the many-body Hilbert space. The state preparation
  is realized by cooperative pumping of nuclear spins by coupled donor
  electrons, and the required controls are {\it in situ} to the
  prototype Kane proposal for quantum computation. This scheme only
  requires a sub-gigahertz donor exchange coupling which can be
  readily achieved without atomically precise donor placement, hence
  it offers a practical way to prepare multipartite entanglement of
  spins in silicon with current technology. All desired Dicke states
  appear as the steady state under various pumping scenarios and
  therefore the preparation is robust and does not require accurate
  temporal controls. Numerical simulations with realistic parameters
  show that Dicke states of $10-20$ qubits can be prepared with high
  fidelity in presence of decoherence and unwanted dynamics.
\end{abstract}
\date{\today}

\pacs{03.67.Bg,76.70.Fz,42.50.Dv,71.55.Cn}


\maketitle

\section{Introduction}
Because of the ultra-long quantum coherence time, electron and nuclear
spins of shallow donors in silicon are ideal candidates as information
carrier in quantum computation. For $^{31}$P donor in isotope purified
$^{28}$Si, a nuclear spin coherence time exceeding 1~s and an electron
spin coherence time exceeding 10~ms were
measured~\cite{morton_solid-state_2008,Tyryshkin_electrondecoherence,Abe_electrondecoherence_2010}.
In his seminal paper~\cite{Kane_QC_nuclei}, Kane proposed an
architecture for scalable quantum computation using nuclear spins of
gated $^{31}$P donors in silicon as qubits. With the superb
compatibility with the existing silicon technology, Kane's proposal
and its variants have stimulated extensive interests in donor systems
in
silicon~\cite{EspinQCSi,Skinner_Kanemodel2,deSousa_QCmodel,Morton_cluster},
and remarkable technological progresses have since been made in
various aspects.  Accurate positioning of the $^{31}$P donor was made
possible by controlled single-ion
implantation~\cite{Jamieson_donorpositioning}.  Local electrical
tuning of the hyperfine interaction between donor electron and nuclear
spins was
demonstrated~\cite{Bradbury_hyperfineTune,Dreher_ElectroelastichyperfineTune}.
An architecture that integrates single $^{31}$P donor with a silicon
SET was recently developed which enables high-sensitivity readout and
control of the donor electron
spin~\cite{morello_architecture_2009,Morello_readout}.  In the
meantime, challenges still remain, e.g. on realizing efficient donor
electron exchange coupling which is also needed for mediating pairwise
interaction of nuclear spins~\cite{DasSarma_exchangeOsc}.  Due to the
interference between the degenerate valleys of the electron, the
exchange coupling strength vastly oscillates with sub-nanometer
variation in donor position (e.g. between 1-100 GHz when donor
separation is $\sim 10$ nm) ~\cite{DasSarma_exchangeOsc}. The chance
to have an exchange coupling of $\geq 30$ GHz expected by the initial
Kane proposal becomes random unless with atomically precise dopant
placement~\cite{Clark_Ppositioning} which is beyond the technological
capability today and in the near future.

In this paper, we introduce a scheme to deterministically prepare Dicke states of donor nuclear spins in silicon. 
Dicke states in general refer to the common eigenstates of $\hat{J}^2$ and $\hat{J}^z$ with eigenvalues $J(J+1)$ and $M$ respectively, where $\hat{\bm J}$ is the collective spin of an ensemble of $N$ spin-$I$ particles. Dicke states with $J$ taking the maximum value $NI$ are symmetric under permutation operations. When $M \neq \pm J$, symmetric Dicke states are a class of genuine multipartite entangled states which have been widely pursued in atomic and optical systems as important resources for quantum information processing~\cite{Haffner_IonDickeState,Zeilinger_PhotonDickeState,Weinfurter_PhotonDickeState,Weinfurter_3qubitWstate,
Duan_AtomDickeState,Sorensen_AtomDickestate,Mabuchi_AtomDickeState,Thiel_AtomPhotonDickeState,Duan_squeezing}. These states have the remarkable properties that the entanglement is robust against qubit loss and projective measurements on the states lead to various entangled states of a lower qubit number.~\cite{Weinfurter_PhotonDickeState} When $J$ takes value other than $NI$, the Dicke states are no longer symmetric under all permutation operations and will be referred here as asymmetric. Asymmetric Dicke states are also resources of genuine multipartite entanglement which are less studied for the lack of preparation schemes. The ideal resource for optimal quantum telecoloning algorithm is one such state with $J=0$.~\cite{Murao} Preparation of asymmetric Dicke states also makes it possible to access the decoherence free subsystems in the presence of collective decoherence~\cite{kempe_theory_2001}. 
The scheme we propose here can deterministically access all Dicke states (symmetric and asymmetric) that span a complete basis for the Hilbert space of $N$ spins. This direct access to the Dicke-states basis can be an important complement to the circuit model quantum information processing, since the entanglement of these collective states can not be achieved in a simple way by pairwise interaction~\cite{Weinfurter_PhotonDickeState}.

Our scheme is based on cooperative pumping of nuclear spins
by the coupled donor electrons, and the required controls are {\it in
  situ} to the prototype Kane proposal: (i) initialization of electron
spin to its ground state in magnetic field; (ii) ac electrical control
of the hyperfine coupling of the donor (A-gate); (iii) on and off
switching of exchange coupling between neighboring electrons (J-gate).
Remarkably, our scheme only needs a sub-gigahertz exchange coupling,
which can be satisfied for almost all donor pairs with separation
$\sim 10$ nm. Hence it provides a practical way for generating the
critical resource of multipartite entanglement of spins in silicon,
which can tolerate the exchange oscillation problem and be realized
within the current technology. Our scheme is a significant example of
the conceptually new approach of dissipative quantum state preparation
with the advantages of robustness and no need for accurate temporal
controls as compared to conventional state preparation by coherent
evolution, as the desired Dicke states all appear as the unique steady
state under the various pumping scenarios. Numerical simulation with
realistic parameters for $^{31}$P donors shows that Dicke states of
$10 - 20$ qubits can be prepared with high fidelity in the presence of
decoherence and unwanted dynamics. The scheme also applies to other
donor systems with larger nuclear spins such as $^{209}$Bi in
silicon~\cite{George_Biqubit,Mohammady_Biqubit,Morley_Biqubit}.

\section{Control scheme}

Fig.~1(a) schematically illustrates the Kane architecture where
shallow donors are embedded under patterned electrodes. The A-gate
tunes the hyperfine interaction $\hat{H}_{\rm hf}=\sum_n a_n\hat{\bm{\sigma}}_n\cdot\hat{\bm{I}}_n $
with $\hat{\bm{\sigma}}_{n}$ and $\hat{\bm{I}}_{n}$ being respectively
the electron spin and nuclear spin of the $n$th donor. The coupling
strength $a_n$ is proportional to the electron density at the donor
nucleus site and hence is a function of the voltage applied to the
A-gate which pulls the electron wavefunction away from the
nucleus. With each A-gate independently controlled with voltage
$V_{n}=V_{n,0}+\delta V_{n}(t)\cos(\omega t+\phi_n)$, the hyperfine
interaction becomes:
\begin{align}
  \hat{H}_{\rm hf} =\sum_{n} a_{n,0}\hat{\bm{\sigma}}_{n}\cdot\hat{\bm{I}}_{n}+\frac{\partial
  a}{ \partial V}\delta V_n\cos(\omega t+\phi_{n})\hat{\bm{\sigma}}_{n}\cdot\hat{\bm{I}}_{n}.
\end{align}
Where the first term is the static hyperfine coupling where $a_{n,0} /
2\pi \equiv a (V_{n,0} ) / 2\pi \simeq 60$ MHz, and the second term is
the ac hyperfine coupling from the voltage modulation. In a strong
magnetic field, the off-diagonal part of the static hyperfine is far
off resonance, hence it only results in small shifts of energy levels
which are neglibile here. However, the off-diagonal part of the ac
hyperfine can efficiently pump nuclear spin polarization when the
modulation frequency $\omega$ is resonant to the electron-nuclear
flip-flop
transition~\cite{rudner_electrically_2007,laird_hyperfine-mediated_2007,Rashba_acDNP}.

The exchange coupling between neighboring donor electrons is tuned by
the J-gate which can separate the electrons. When the electrons are decoupled, nuclear spins of different donors can be independently pumped to the fully polarized state (i.e. nuclear spin
initialization). When the electrons are coupled, nuclear spins can be
cooperatively flipped by collective raising/lowering operators. We
have previously proposed using such pumping to probabilistic prepare
singlets of nuclear spins~\cite{yao_many-body_2011,Yu_SCS_2011}.

\begin{figure}[t]
\includegraphics[width=8cm]{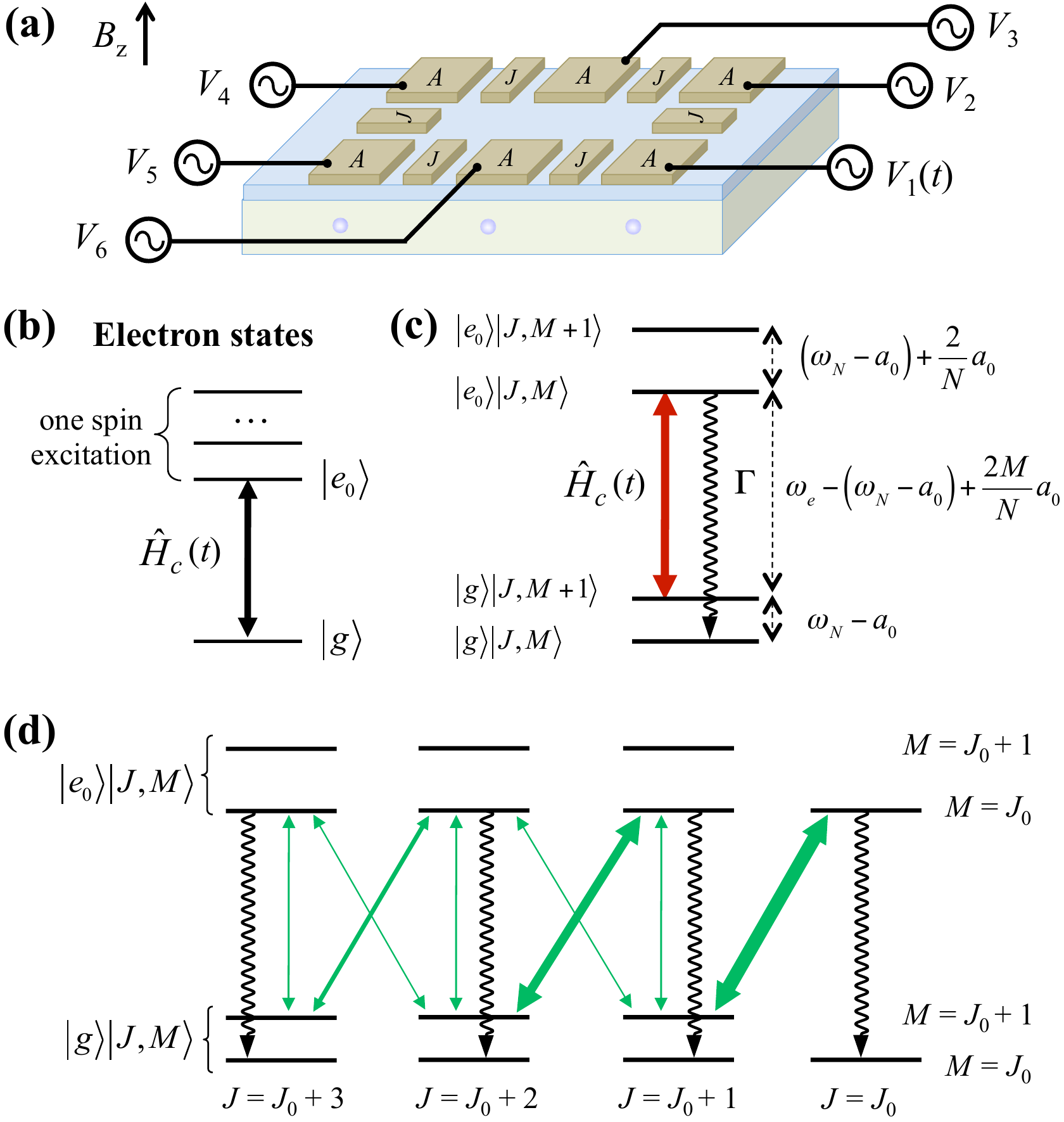}
\caption{(a) Donor spins in silicon controlled by patterned
  electrodes. Pumping of nuclear spins is realized through ac voltage
  control of the A gates. (b) Energy level scheme for the exchange
  coupled electrons in magnetic field along $z$ direction. The
  transition between the ground state $|g\rangle$ and a non-degenerate
  one-spin excitation state $|e_0\rangle$ is used. (c) and (d) Pumping
  of nuclear spins by ac hyperfine coupling $\hat{H}_c$ (double head
  arrows), assisted by the electron spin decay channel from
  $|e_0\rangle$ to $|g \rangle$ (wavy lines). The central frequency of
  $\hat{H}_c$ is tuned in resonance with the energy separation between
  $|g \rangle |J, M+1 \rangle$ and $ | e_0 \rangle | J', M \rangle$
  which is a quantity dependent on $M$. (c) When the ac voltage
  control uses the first scenario for phases (see text), population is
  deterministically transferred from the collective nuclear state $|J,
  M+1 \rangle$ to $| J, M \rangle$. (d) When the ac voltage control
  uses the second scenario for phases (see text), population is
  transfered from $|J, M+1 \rangle$ to $| J', M \rangle$ where
  $J'-J=0$ or $\pm 1$.} \label{fig1}
\end{figure}

In general, for coupled donor electrons in uniform magnetic field
along $z$ direction, one can always find a non-degenerate
one-spin-excitation eigenstate $|e_{0}\rangle \equiv
\sum_{n}\alpha_{n}\hat{\sigma}_{n}^{+}|g\rangle$, with $|g\rangle
\equiv |\downarrow \cdots \downarrow \rangle$ denoting the ground
state.  With the ac control frequency $\omega$ tuned near resonance
with the transition between $|g\rangle$ and $|e_{0} \rangle$, we can
neglect all other far detuned electron spin resonances. Dropping
non-secular terms, the full Hamiltonian of the electron nuclear spin
system can then be written as $\hat{H} = \hat{H}_0 + \hat{H}_c$ where:
\begin{subequations}
\begin{eqnarray}
  \hat{H}_0 &=& \omega_e |e_0 \rangle \langle e_0 | + \omega_N \sum_n  \hat{I}^z_n \label{Hdiag} \\ 
  && - |g \rangle \langle g | \sum_n a_{0}\hat{I}^z_n - \frac{N-2}{N} |e_0 \rangle \langle e_0 | \sum_n a_{0} \hat{I}^z_n  ~, \notag \\
  \hat{H}_c & = & e^{i\omega t} |e_0\rangle \langle g| \sum_{n}  \Omega_n (t) e^{i\phi_{n}}\hat{I}^-_{n} + h.c. ~.  \label{Hoffdiag}
\end{eqnarray}
\end{subequations}
$\omega_e$ is the electron resonance frequency between $|e_0\rangle$
and $|g\rangle$ in the magnetic field, $\omega_N$ is the nuclear
Zeeman frequency, and $\Omega_n \equiv \frac{\partial a}{ \partial V}
\delta V_n \alpha_{n} $. The dc voltage of each A gate is set such that $a_{n,0}=a_0$. $\hat{H}_c$ induces electron-nuclear flip-flop where the nuclear spin flip is in a cooperative form
determined by the phases $\phi_n$ and amplitudes $\delta V_n$ of the
ac controls. For simplicity, we consider hereafter a uniformly coupled Heisenberg ring with the eigenstate $|e_{0} \rangle = \sum _n \frac{1}{\sqrt{N}} (-1)^n\hat{\sigma}_{n}^{+}|g\rangle$
which is gapped from other one-spin-excitation states by $\Delta$
(Fig.~1(b)). Then we have $\Omega_n (t) = (-1)^n \Omega(t) $ where $\Omega \equiv \frac{1}{\sqrt{N}} \frac{\partial a}{ \partial V}
\delta V $.

We consider two scenarios for the phases $\phi_n$. In the first
scenario, $(-)^n e^{i\phi_{n}}=1$ for all donors and hence the
electron-nuclear flip-flop term becomes: $ \hat{H}_c = \Omega (t)
[e^{i \omega t} |e_0 \rangle \langle g | \hat{J}^- + h.c.]$. Here
$\hat{J}^{\pm} = \hat{J}^{x} \pm i \hat{J}^y$ where $\hat{\bm J}
\equiv \sum_n \hat{\bm I}_n$ is the collective spin of all nuclei. In
the second scenario, $(-)^n e^{i\phi_{n}}=1$ for a set of donors
(referred as group A) while $(-)^n e^{i\phi_{n}}=-1$ for the rest
(referred as group B), and hence the electron-nuclear flip-flop term
is of the form: $\hat{H}_c = \Omega (t) [ e^{i \omega t} |e_0 \rangle
\langle g | (\hat{j}_A^- - \hat{j}_B^-) + h.c.]$, where $\hat{\bm
  j}_A$ and $\hat{\bm j}_B$ are the collective spin of nuclei in group
A and group B respectively, and $ \hat{\bm j}_A + \hat{\bm j}_B =
\hat{\bm J}$.  Dicke states here refer to the common eigenstates of
$\hat{J}^2$ and $\hat{J}^z$ with eigenvalues $J(J+1)$ and $M$
respectively. Consider first the subspace with $j_A=n_A I$ and
$j_B=(N-n_A) I$ where $n_A$ and $N-n_A$ are the number of nuclei (of
spin-$I$) in group A and B respectively. $j_A$ and $j_B$ are conserved
quantum numbers in the dynamics. Dicke states in this subspace can be
uniquely specified as $|J, M \rangle$.

\begin{figure}[t]
\includegraphics[width=8cm]{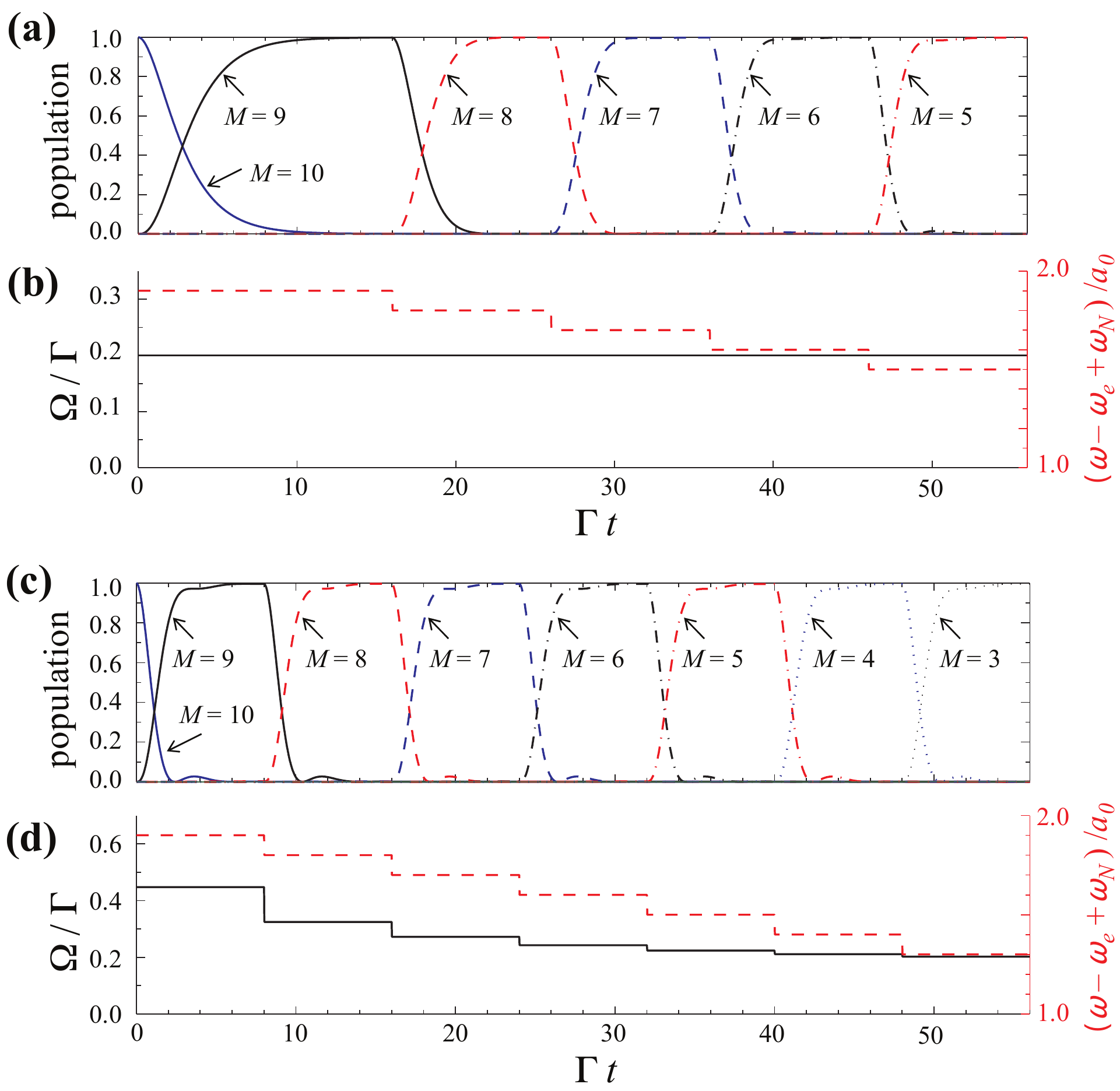}
\caption{(Color online) Preparation of the symmetric Dicke states of
  20 nuclear spin qubits initially on the fully polarized state. For
  the numerical simulation in (a-b), the amplitude $\Omega$ of the ac
  control is held as a constant while the central frequency $\omega$
  steps down after a finite interval. Symmetric Dicke states $|J=10,M
  \rangle$ with different $M$ are obtained sequentially with a
  probability $\geq 99.8 \%$. In (c-d), the control sequence
  $\Omega(t)$ is optimized for speed, and each Dicke state is obtained
  with a probability $\geq 99.5 \%$. $\Gamma/ 2
  \pi=60$~kHz.} \label{fig2}
\end{figure}

The first non-trivial element for preparing Dicke states is the
deterministic population transfer from $|J, M+1\rangle$ to $|J,
M\rangle$. Only the first scenario for the phase control is needed
where $\hat{H}_c$ conserves the quantum number $J$.  We assume the
population on electron excited state $|e_0\rangle$ can be efficiently
dumped to ground state with rate $\Gamma$, which can be realized,
e.g. via the tunneling process between the donor and the SET island at
low temperature~\cite{morello_architecture_2009,Morello_readout}.  If
the ac control frequency $\omega$ is in resonance with the transition
$|g\rangle |J, M+1\rangle \leftrightarrow |e_0\rangle |J, M \rangle$,
the population on $|g\rangle |J, M+1 \rangle$ is pumped one-way to
$|g\rangle |J, M \rangle$ via the Raman-type process [Fig.~1(c)]. For
efficient initialization of the electron spin on the ground state, the
electron Zeeman energy shall be large as compared to the
temperature. With a typical temperature of $\sim 100$ mK used in the
experiments~\cite{Morello_readout}, a magnetic field $\geq 0.3$ T is
required, which corresponds to ac modulation of the A-gate voltage in
the frequency range $\geq 10$ GHz.

There exists two unwanted couplings by the electron-nuclear flip-flop:
(1) the coupling to other electron one-spin-excitation states which
are detuned at least by $\Delta$ [Fig.~1(b)]; (2) the out-coupling of
the final state $|g\rangle |J, M\rangle $ to $ |e_0\rangle |J, M-1
\rangle$ which is also detuned by $\frac{2}{N} a_0$. For the latter,
we note that for any pair of states $|g\rangle |J, M+1\rangle$ and
$|e_0\rangle |J', M\rangle$ coupled by $\hat{H}_c$, the resonant
frequency is $\omega_e - \omega_N+a_0 + \frac{2M}{N} a_0$, namely, the
pumping is $M$-selective with proper choice of the ac control
frequency. When the detuning $\delta \equiv \min \{\Delta, \frac{2}{N}
a_0\}\gg \Gamma$, these unwanted couplings cause a negligible leakage
$\propto (\frac{\Gamma}{\delta})^2$.

\begin{figure}[t]
\includegraphics[width=8cm]{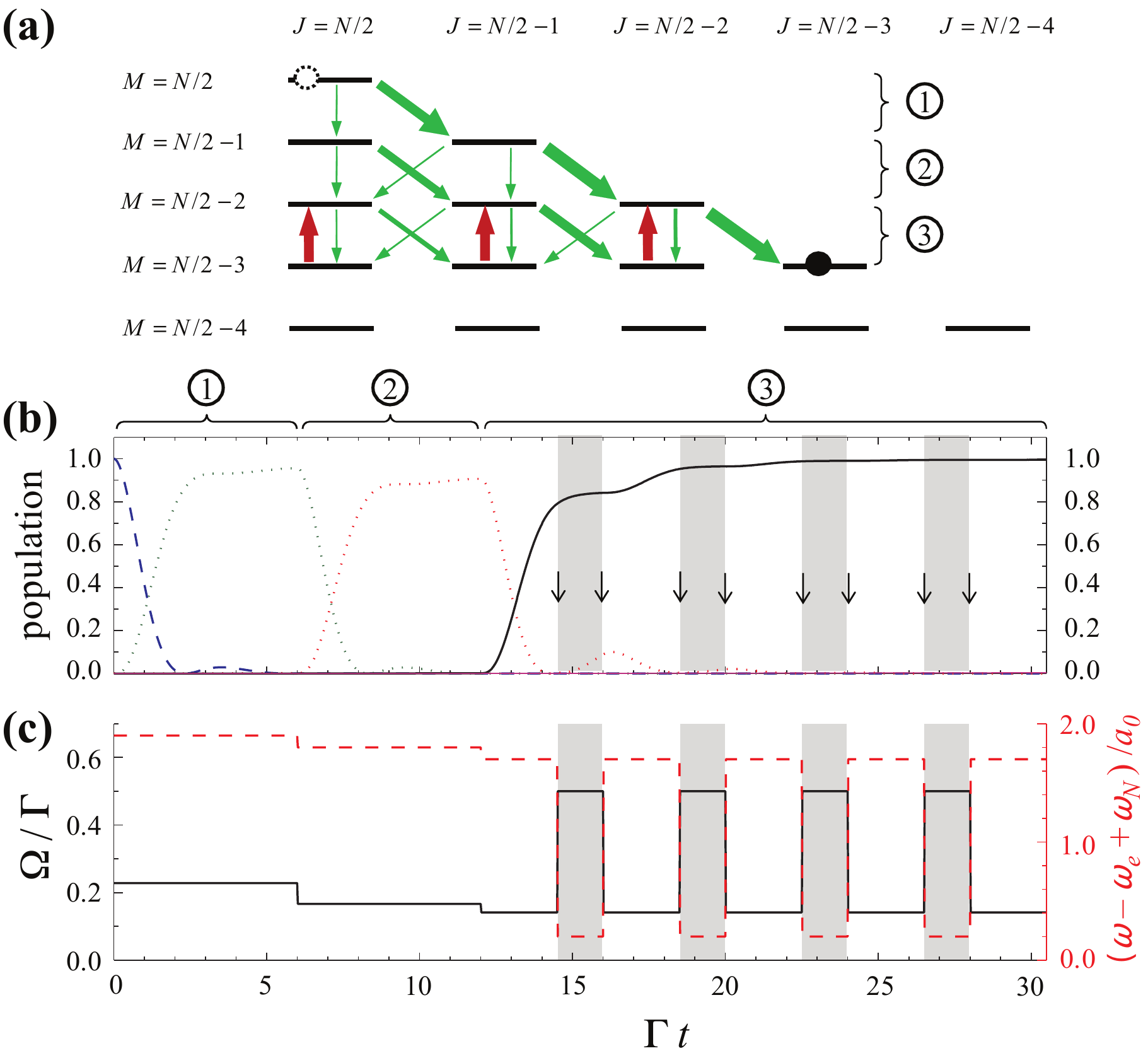}
\caption{(Color online) Preparation of asymmetric Dicke state $|J = 7,
  M=7 \rangle$ in the subspace $(j_A=4, j_B=6)$ for 20 nuclear spin
  qubits initially on the fully polarized state. (a) Population flows
  by the cooperative pumping.  (b) and (c) show numerical simulation
  of the state preparation. (b) Populations on the various Dicke
  states: $|J = 10, M=10\rangle$ (dashed blue); $|J = 9, M=9 \rangle$
  (dotted green); $|J = 8, M=8\rangle$ (dotted red). The solid curve
  gives the population on the target state $|J = 7, M=7 \rangle$ (or
  its time reversal state in the shaded interval), and its final
  population is $99.5\%$. (c) The magnitude $\Omega$ and central
  frequency $\omega$ for the ac tuning of hyperfine interaction. In
  the unshaded (shaded) interval, the first (second) scenario for the
  phases is applied (see text). The vertical arrows indicate timing of
  global $\pi$ flip of all nuclear spins. $\Gamma/ 2
  \pi=60$~kHz. } \label{fig3}
\end{figure}

With the nuclear spins initialized on the unentangled polarized state
$|J=N I, M=N I \rangle$, the above simple control can
deterministically prepare all symmetric Dicke states. Examples for
preparing symmetric Dicke states of $N=20$ $^{31}$P nuclear spins are
simulated using the master equation
\begin{align}
  \dot{\rho} = i [\rho, \hat{H}] - \frac{\Gamma}{2} \left( |e_0
    \rangle \langle e_0 | \rho + \rho |e_0 \rangle \langle e_0 | -2
    |g\rangle \langle e_0| \rho |e_0\rangle \langle g|\right).
  \label{me}
\end{align}
The results are shown in Fig.~2. Clearly, the state preparation has
the advantage that it is insensitive to the shape and area of the
control pulse $\Omega(t)$. We take $a_0/ 2 \pi=60$~MHz, $\Gamma/ 2
\pi=60$ kHz, and assume the donor electron exchange of 0.5 GHz which
leads to $\Delta/ 2 \pi \sim 25$ MHz.  The preparation takes a time
$t_p \sim O(10)~\mu$s. In comparison, a single two-qubit nuclear spin
gate mediated by such a small donor exchange would take $\sim 1$
ms~\cite{Hill_Gate}.  The probability to obtain each Dicke state is
nearly unit ($\geq 99.5 \%$), and the imperfection is caused by the
off-resonance couplings which maybe further reduced by using smaller
$\Omega$ with the cost of longer preparation time.  We further note
that the reverse population transfer from $|J, M \rangle$ to $|J,
M+1\rangle$ can be realized if global $\pi$ flips of nuclear spins are
applied before and after the above pumping process with the central
frequency set at $\omega=\omega_e - \omega_N +a_0 - \frac{2(M+1)}{N} a_0$.

The second non-trivial element is to realize the asymmetric Dicke
state $|J=J_0, M=J_0\rangle$ with general values of $J_0$. This
requires the quantum number $J$ to be changed which needs the control
with the second scenario for phases. As discovered by the authors in
our earlier work~\cite{yao_many-body_2011,Yu_SCS_2011}, the
inhomogeneous collective operator $\hat{j}_A^{-} - \hat{j}_B^{-}$
couples $|J, M+1\rangle$ to the states $|J+\Delta J, M\rangle$ with
the selection rule $\Delta J = 0, \pm 1$. Hence $\hat{H}_c$ can
resonantly drive the one-way pumping from $|J, M+1\rangle$ to
$|J+\Delta J, M\rangle$ (see Fig.~1(d)). By repeating the above
pumping for $NI-J_0$ steps, the population can be transferred from the
initial state $|J=N I, M=N I\rangle$ to the states $|J\geq J_0,
M=J_0\rangle$ (see Fig.~3(a)). If the reverse pumping from $|J,
M=J_0\rangle$ to $|J, M=J_0 +1\rangle$ (see proceeding paragraph) is
also turned on, then all pathways from the initial state will end up
at the desired target state $|J=J_0, M=J_0\rangle$ where the
population gets trapped (see Fig.~3(a)). This realizes the
deterministic preparation of $|J=J_0, M=J_0\rangle$ with arbitrarily
specified $J_0$. Fig.~3(b) shows the numerical simulation for
preparing the Dicke state $|J = 7, M=7, j_A=4, j_B=6 \rangle$ of 20
$^{31}$P nuclear spins, obtained with a probability of $99.5\%$.

\begin{figure}[t]
\includegraphics[width=7cm]{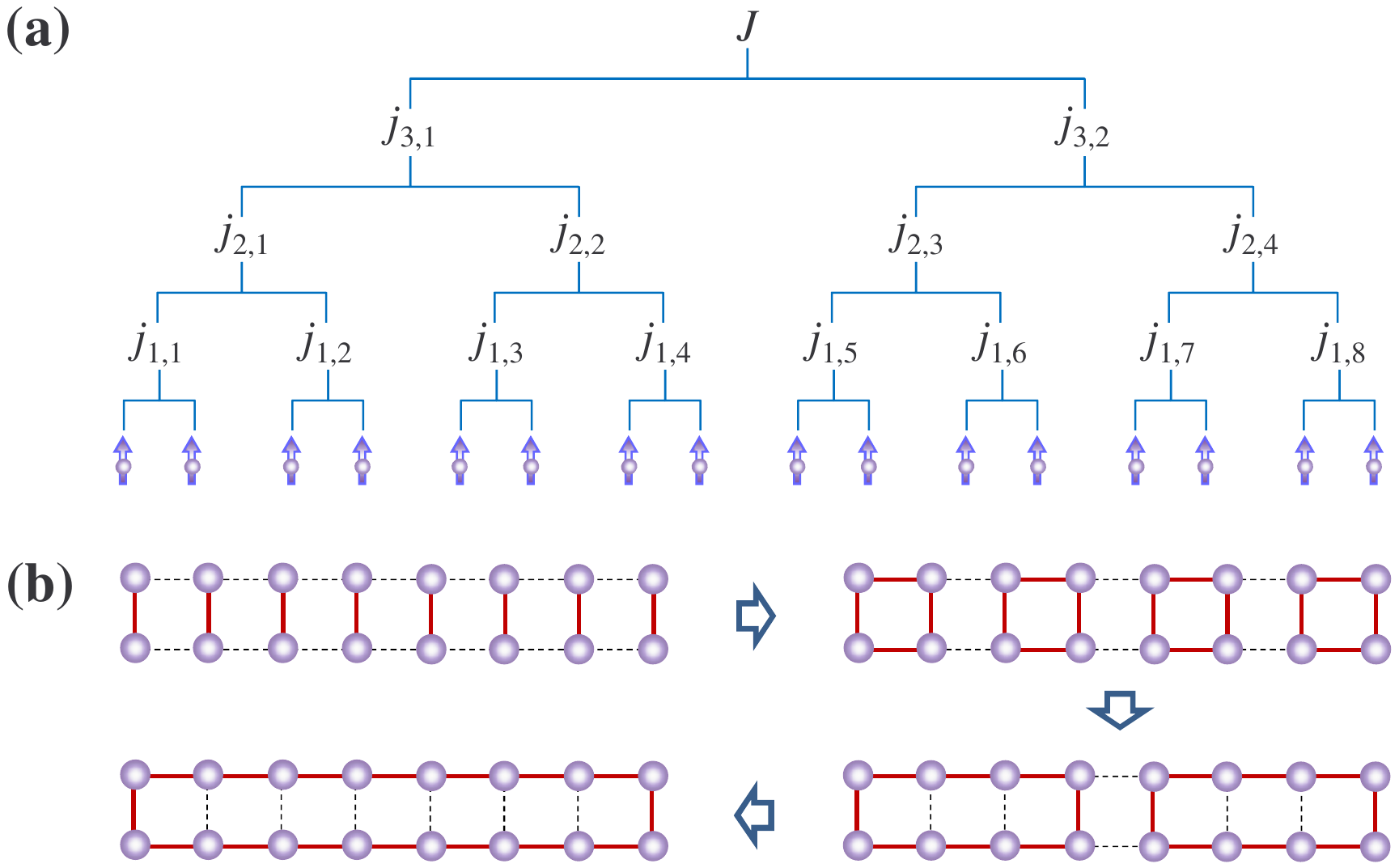}
\caption{Preparation of Dicke states $|J,M,\{j_{l,k} \} \rangle$ which
  spans a complete basis for the nuclear spin qubits. (a) The quantum
  numbers $j_{l,k}$ are the collective spin of subsets of qubits. (b)
  Concatenated preparation process. In step $l$, Dicke states are
  parallelly prepared in each $2^l$-qubit subset with the collective
  spin being the specified value $j_{l,k}$. The quantum numbers
  $j_{l<l_0,k}$ are all conserved in step $l_0$. Red solid (black
  dashed) lines indicate the donor electron exchange `on' (`off') when
  nuclear spins are pumped.} \label{fig4}
\end{figure}

Combining these two elements, an arbitrary Dicke state can be
deterministically prepared in the subspace defined by $j_A= n_A I$ and
$j_B=(N-n_A) I$. Essentially, this is the realization of an arbitrary
total spin eigenstate for the two collective spins $\hat{\bm j}_A$ and
$\hat{\bm j}_B$. By concatenating this procedure, we can
deterministically prepare any state in a complete Dicke states basis
for $N$ spins, denoted as $|J, M, \{j_{l,k} \} \rangle $ where
$\{j_{l,k} \}$ are the collective spin of subsets [Fig.~4(a)]. The
concatenated procedures are illustrated in Fig.~4(b).

\section{Effects of possible imperfections}

Here we analyze effects of various imperfections that may exist in realistic systems and give recipes on how to deal with these sources of errors. 

The first cause of error is the decoherence of the spin qubits. Dicke states are immune to decoherence in a collective form, but independent nuclear spin decoherence can cause a leakage out of the desired subspace. For a $N$-qubit state, the leakage is $\sim t_p N \gamma$ where $\gamma$ is the nuclear spin decoherence rate and $t_p$ is the preparation timescale proportional to $ \Gamma^{-1}$. The longest nuclear spin coherence time reported for $^{31}$P donor in silicon is $1.75$ second.~\cite{morton_solid-state_2008} In Table.~\ref{table1}, we show the performance of preparing several exemplary Dicke states in presence of spin decoherence. Nuclear spin pure dephasing described by the Lindblad term $- \frac{\gamma}{2} \sum_n \left( \hat{I}^z_n \hat{I}^z_n \rho + \rho \hat{I}^z_n \hat{I}^z_n - 2\hat{I}^z_n \rho \hat{I}^z_n \right)$ is added to the master equation. The much slower nuclear spin relaxation process is neglected~\cite{mccamey_electronic_2010}. Because of the short preparation timescale, there is no visible effect from the decoherence if we take $\gamma / 2 \pi=0.1$ Hz from the state-of-art measurement~\cite{morton_solid-state_2008}. Even with the much exaggerated decoherence rate $\gamma/ 2 \pi=10$ Hz, the target states can still be obtained with high fidelity.

\begin{table}[t]
\caption{Figure of merits for preparing Dicke states of 10 qubits in presence of decoherence. $|\psi_1\rangle \equiv |J=5,M=4\rangle$ is the W-state. $|\psi_2\rangle\equiv |J=5,M=0\rangle$ is the symmetric Dicke state with the most spin excitations. $|\psi_3\rangle\equiv |J=2,M=2,j_A=\frac{5}{2},j_B=\frac{5}{2}\rangle$ is an asymmetric Dicke state. $F\equiv \langle \psi | \rho | \psi \rangle $ is the fidelity of the density matrix with the target state at $t=t_p$. We take $\Gamma/ 2 \pi=50$ kHz and $\Omega_0$ is the peak value of $\Omega(t)$ used.} \label{table1}
\begin{ruledtabular}
\begin{tabular}{cccccc}
target 		 & $\Omega_0 / 2 \pi$  &  $t_p$       	   & $F$			 	  & $F$   \\
 			 &              	            &  		     	   & ($\gamma/ 2\pi$ = 0.1 Hz)    & ($\gamma/ 2 \pi$ = 10 Hz)   \\ \hline
$|\psi_1\rangle$ & 31.5 kHz	   	    & 32 $\mu$s    & 0.999 				  & 0.998  \\
$|\psi_2\rangle$ & 31.5 kHz	   	    & 108 $\mu$s  & 0.999 				  & 0.987 \\
$|\psi_3\rangle$ & 41 kHz	   	   	    & 175 $\mu$s  & 0.998 				  & 0.982 \\
\end{tabular}
\end{ruledtabular}
\end{table}

In realistic systems, there also exist various defects in the surrounding of the donors such as the interface P$_{b0}$ centers.~\cite{Dreher_ElectroelastichyperfineTune,Stegner_Pb0_2006,Huebl_Pb0_2008,Morley_Pb0_2008,Morishita_Pb0_2009}
The interplay between P$_{b0}$ centers and phosphorus electron spins is important in the presence of photo-excited electrons and holes. Such interplay can facilitates the electrical detection of paramagnetic resonance of the donor electron~\cite{Dreher_ElectroelastichyperfineTune,Stegner_Pb0_2006,Huebl_Pb0_2008,Morley_Pb0_2008,Morishita_Pb0_2009}. In the absence of illumination as in our scenario, at
temperature of $k_B T\ll g\mu B$, the P$_{b0}$ center as a mid-gap paramagnetic center can lead to statistical fluctuations of the electron-nuclear flip-flop resonances from donor to donor. Furthermore, there could be noises from the local electrostatic environment, e.g. possible charge traps in the gate oxide~\cite{Morello_readout}, which may affect the hyperfine coupling and the donor electron exchange as well. If the charge hopping is faster than the state preparation timescale, the effect of this dynamics noise shall be similar to the effect of spin decoherence analyzed above. It is also possible to interlace the pumping control with sequence of $\pi$ pulses applied to the spin qubits for dynamical decoupling from such noises.~\cite{GongZR} If charge hopping is much slower as compared to the state preparation timescale, the noises are static which also result in the statistical fluctuation of the hyperfine coupling strength and exchange coupling strength.

To deal with such statistical fluctuations of system parameters, the system provides multi-fold individual tunability for each donor. The electron-nuclear flip-flop resonance depends on the static part of the hyperfine interaction (see Eq.~(2a)) which can be independently controlled by each A-gate. Thus, in presence of inhomogeneous broadening of the electron-nuclear flip-flop resonances, the dc part of the A-gate voltage can tune the static part of the hyperfine coupling for compensating the inhomogeneity after calibration. As the excited state for the electron-nuclear flip-flop has an intrinsic broadening of $\Gamma$, inhomogeneity in the flip-flop resonances is unimportant if it is less than this intrinsic level broadening. 

The exchange couplings between donor electrons are in general inhomogeneous due to the exchange oscillation problem and the possible effects of the aforementioned charge noises. This inhomogeneity manifests as nonuniform amplitudes $\alpha_n$ in the electron excited state $|e_{0}\rangle \equiv \sum_{n}\alpha_{n}\hat{\sigma}_{n}^{+}|g\rangle$.
We note that the pumping operators that cooperatively flips the nuclear spin qubits are of the form: $\sum_{n}  \frac{\partial  a}{ \partial V} \delta V_n  \alpha_{n} e^{i\phi_{n}}\hat{I}^-_{n}$ (see Eq.~(\ref{Hoffdiag})). While J-gate voltages can tune the exchange couplings and hence the values of $\alpha_n$, the ac modulation strength $\delta V_n$ and phase $\phi_n$ can further compensate the remaining inhomogeneity in $\alpha_n $. 

If only entanglement generation is of interest, our scheme can naturally cope with unknown systematic errors in the cooperative pumping. For controls aimed at the symmetric Dicke state $(\hat{J}^-)^m|J=N/2, M=N/2\rangle$, the nuclear spins may be flipped instead by $\hat{A}^- \equiv \sum_n (1+\eta_n)e^{i \theta_n} \hat{I}_n^-$ where $\theta_n$ and $\eta_n$ are respectively the unknown phase and amplitude errors. In presence of these errors, the steady state is still a definite pure state $(\hat{A}^-)^m|J=N/2, M=N/2\rangle$ with similar multipartite entanglement. This feature makes the requirement less stringent for implementing this scheme for entanglement generation. 

Moreover, one advantage of our scheme is unwanted donor sites can be easily disconnected in the state preparation. There are two ways to do this. One is to turn off the coupling of such donor site with others by the J-gate control. Even if the exchange interaction couples all neighboring donor electron spins into collective states, the set of nuclear spins being pumped can be further selected by the ac voltage controls. If the voltage applied to one of these donors has no ac component, its nuclear spin will not be pumped and is effectively decoupled from the rest in the preparation process. This advantage allows one to select out the set of donors where the parameters are relatively uniform, and it also facilitates the realization of the concatenated preparation process described in Fig.~4.

Finally, we give an estimate on the scale of the Dicke states that can be prepared using this approach. First, the detuning of the unwanted transitions is $\sim\frac{2}{N}a_0$ which
scales inversely with the qubit number $N$. To suppress the unwanted transitions, we
require $\frac{2}{N}a_0\gg\Gamma$. Second, the leakage due to qubit decoherence is $\sim t_p N \gamma$, while the preparation time $t_p$ is determined by the
smaller one of the modulation amplitude of the hyperfine interaction
$\Omega$ and the spin initialization rate $\Gamma$. In our simulation, we take $\Gamma \sim50-60$ kHz and $\Omega \sim30-40$ kHz, both values in the range
reported in experiments \cite{Dreher_ElectroelastichyperfineTune,
  Bradbury_hyperfineTune, Morello_readout}. Then $N$ shall not exceed $100$ by the first requirement mentioned above. The upper bound for $N$ set by the second requirement depends on the qubit decoherence rate $\gamma$. For given $\gamma$, the
fidelity for state preparation in a block of $N$ qubits may be scaled
from the numbers listed in Table I. For example, with the nuclear spin decoherence rate of $10$ Hz ($T_2\sim15$ ms), preparation of asymmetric Dicke state $|\psi_3\rangle$ for $20$ qubits
can have a fidelity over $0.96$. Symmetric Dicke states may be prepared in a much larger scale, since they take much less steps by the cooperative pumping and $t_p$ is then considerably shorter. 

The work was supported by the Research Grant Council of Hong Kong under Grant No. HKU706711P and HKU8/CRF/11G. The authors acknowledge Andrea Morello and Wendi Li for helpful discussions.

\end{document}